\documentclass[conference, a4paper]{IEEEtran}

\usepackage{tabularx} 

\ifCLASSINFOpdf
\else
\fi
\usepackage{amsfonts,amsmath,amssymb}
\usepackage{mathtools}
\usepackage{mdwmath}
\usepackage{cuted}
\usepackage{mdwtab}
\usepackage{amsthm}           
\usepackage{bm}               
\usepackage{units}            
\usepackage{graphicx}
\usepackage{epstopdf}
\usepackage[linesnumbered,ruled,vlined]{algorithm2e}
\usepackage{comment}

\usepackage{url}

\usepackage{paralist}
\usepackage{color}
\usepackage{authblk}
\usepackage{cleveref}
\usepackage{float}
\usepackage[caption = false]{subfig}
\usepackage[nolist,nohyperlinks]{acronym} 

\newcommand{\Fig}[1]{Fig.~\ref{fig:#1}}

\newcommand{\Eq}[1]{Eq.~(\ref{eq:#1})}

\newcommand{\Rb}{\mathbb{R}}

\newcommand{\sinr}{\mathrm{SINR}}


\begin{acronym}[NC-OFDM]
\acro{3gpp}[3GPP]{3\textsuperscript{rd} Generation Partnership Program}
\acro{5g}[5G]{Fifth Generation}
\acro{Adam}{Adaptive Moment Optimisation}
\acro{anr}[ANR]{Automatic Neighbour Relation}
\acro{ap}[AP]{Access Point}
\acro{atc}[ATC]{Air Traffic Control}
\acro{bs}[BS]{Base Station}
\acro{bpp}[BPP]{Binomial Point Process}
\acro{cc}[C\&C]{Command \& Control}
\acro{cdf}[CDF]{cumulative distribution function}
\acro{cdma}[CDMA]{Code Division Multiple Access}
\acro{cdr}[CD\&R]{Conflict Detection \& Resolution}
\acro{cfo}[CFO]{Carrier Frequency Offset}
\acro{comreg}[ComReg]{Commission for Communications Regulation}
\acro{cho}[CHO]{Conditional Handover}
\acro{dqn}[DQN]{Deep Q-Network}
\acro{ddqn}[DDQN]{Dueling Deep Q-Network}
\acro{csi}[CSI]{Channel State Information}
\acro{easa}[EASA]{European Aviation Safety Agency}
\acro{ee}[EE]{Energy Efficiency}
\acro{faa}[FAA]{Federal Aviation Administration}
\acro{fso}[FSO]{Free-Space Optical}
\acro{geo}[GEO]{Geosynchronous Equatorial Orbit}
\acro{gps}[GPS]{Global Positioning System}
\acro{gs}[GS]{Ground Station}
\acro{hap}[HAP]{High Altitude Platform}
\acro{iot}[IoT]{Internet of Things}
\acro{ipnn}[IPNN]{Interference Prediction Neural Network}
\acro{kpi}[KPI]{Key Performance Indicator}
\acro{lap}[LAP]{Low Altitude Platform}
\acro{leo}[LEO]{Low Earth Orbit}
\acro{los}[LoS]{Line-of-Sight}
\acro{lte}[LTE]{Long Term Evolution}
\acro{lstm}[LSTM]{Long Short-term Memory}
\acro{mac}[MAC]{Media Access Control}
\acro{mcp}[MCP]{Matern Cluster Process}
\acro{mc}[MC]{Monte Carlo}
\acro{3gpp}[3GPP]{3rd generation partnership project}
\acro{mimo}[MIMO]{Multiple Input Multiple Output}
\acro{mip}[MIP]{Mixed-Integer Programming}
\acro{mm}[MM]{Mapping Mechanism}
\acro{ml}[ML]{machine learning}
\acro{mno}[MNO]{Mobile Network Operator}
\acro{milp}[MILP]{Mixed-Integer Linear Programming}
\acro{nlos}[NLoS]{non-Line-of-Sight}
\acro{nn}[NN]{Neural Network}
\acro{nrt}[NRT]{Neighbour Relation Table}
\acro{ofdma}[OFDMA]{Orthogonal Frequency Division Multiple Access}
\acro{oot}[OOT]{Out-of-Tree}
\acro{osi}[OSI]{Open Systems Interconnection}
\acro{otdoa}[OTDoA]{Observed Time Difference of Arrival}
\acro{ott}[OTT]{Over-The-Top}
\acro{pv}[PV]{photo-voltaic}
\acro{pdf}[pdf]{probability density function}
\acro{ppp}[PPP]{Poisson Point Process}
\acro{qos}[QoS]{Quality of Service}
\acro{rc}[RC]{Remote Control}
\acro{rl}[RL]{Reinforcement Learning}
\acro{reqiba}[REQIBA]{REgression and deep Q-learning for Intelligent UAV cellular user to Base station Association}
\acro{rss}[RSS]{Received Signal Strength}
\acro{se}[SE]{Spectral Efficiency}
\acro{sir}[SIR]{Signal-to-Interference Ratio}
\acro{sinr}[SINR]{Signal-to-Interference-and-Noise Ratio}
\acro{snr}[SNR]{Signal-to-Noise Ratio}
\acro{uav}[UAV]{Unmanned Aerial Vehicle}
\acro{ue}[UE]{User Equipment}
\acro{ula}[ULA]{Uniform Linear Array}
\acro{wsn}[WSN]{Wireless Sensor Network}
\acro{wsn}[WSN]{Wireless Sensor Network}
\acro{rv}[RV]{random variable}
\acro{ppp}[PPP]{Poisson point process}
\acro{pgfl}[PGFL]{point generation functional}
\acro{pdf}[PDF]{probability density function}
\end{acronym}

\begin{document}
%
\title{Multi-Agent Deep Reinforcement Learning For Optimising Energy Efficiency of Fixed-Wing UAV Cellular Access Points}
%
%
%

\author{Boris Galkin,
        Babatunji Omoniwa,
        and Ivana Dusparic
}

\affil{CONNECT- Trinity College Dublin, Ireland \\
\textit{E-mail: \{galkinb, omoniwab, duspari\}@tcd.ie}}

\maketitle

\begin{abstract}
\acp{uav} promise to become an intrinsic part of next generation communications, as they can be deployed to provide wireless connectivity to ground users to supplement existing terrestrial networks. The majority of the existing research into the use of \ac{uav} access points for cellular coverage considers rotary-wing \ac{uav} designs (i.e. quadcopters). However, we expect fixed-wing \acp{uav} to be more appropriate for connectivity purposes in scenarios where long flight times are necessary (such as for rural coverage), as fixed-wing \acp{uav} rely on a more energy-efficient form of flight when compared to the rotary-wing design. As fixed-wing \acp{uav} are typically incapable of hovering in place, their deployment optimisation involves optimising their individual flight trajectories in a way that allows them to deliver high quality service to the ground users in an energy-efficient manner. In this paper, we propose a multi-agent deep reinforcement learning approach to optimise the energy efficiency of fixed-wing \ac{uav} cellular access points while still allowing them to deliver high-quality service to users on the ground. In our decentralized approach, each \ac{uav} is equipped with a \ac{ddqn} agent which can adjust the 3D trajectory of the \ac{uav} over a series of timesteps. By coordinating with their neighbours, the \acp{uav} adjust their individual flight trajectories in a manner that optimises the total system energy efficiency. We benchmark the performance of our approach against a series of heuristic trajectory planning strategies, and demonstrate that our method can improve the system energy efficiency by as much as 70\%. 
\end{abstract}

\begin{IEEEkeywords}
Cellular-connected UAVs, deep reinforcement learning, energy efficiency.
\end{IEEEkeywords}

\acresetall
\section{Introduction}
\label{sec:Introduction}
\ac{uav} access points are becoming an attractive option for delivering enhanced wireless coverage in cellular networks \cite{Mozaffari2019_UAV_Survey}. Specifically, \acp{uav} can provide additional capacity to service demand hotspot areas and deliver network coverage in isolated or hard-to-reach rural areas where fixed terrestrial infrastructure may be unavailable or insufficient \cite{Mozaffari2019_UAV_Survey}. \ac{uav} infrastructure offers several important advantages to network operators, in that \acp{uav} can be rapidly deployed to provide connectivity to ground users, and leverage their adjustable altitude, obstacle-avoidance capabilities, and strong \ac{los} communication links to maximise performance.


Despite a growing interest in \acp{uav}, several technical challenges must be addressed to effectively utilise them for each specific networking application. These challenges include optimising 3D deployment and placement~\cite{Liu2019UAVqlearning},~\cite{omoniwa_2021}, \ac{ee}~\cite{Liu2019UAVqlearning},~\cite{omoniwa_2021},~\cite{Kuo_2021}, and flight trajectory \cite{Visintini_2021}. Compared to rotary-wing \acp{uav} (such as quadcopters), the fixed-wing \acp{uav} have a better energy-efficient mode of flight due to their mechanical design \cite{Zeng_2017}, therefore we anticipate their use for connectivity purposes in scenarios  where long  flight  duration is required, such as for rural coverage. Nevertheless, designing the \acp{uav} to perform their connectivity tasks over extended periods of time is a significant challenge, due to their limited on-board battery energy. At the time of writing, there has been some research into the energy management of fixed-wing \ac{uav} systems in relay~\cite{Song_2020},~\cite{Ahmed_2020} and data aggregation~\cite{Kuo_2021} applications; however, there is a significant lack of research which investigates fixed-wing \acp{uav} acting as flying base stations for delivering coverage.

The work in~\cite{Song_2020} and \cite{Ahmed_2020} propose conventional optimisation methods to maximise the \ac{ee} of a fixed-wing \ac{uav}. However, they only consider a single fixed-wing \ac{uav} relay to amplify and forward (AF) received signals between source(s) and destination. In geographically-large rural areas with scattered concentrations of users, multiple fixed-wing \acp{uav} may be deployed~\cite{Kuo_2021}, \cite{PeterHong_2020}, \cite{Shriwastav_2020}. In \cite{Kuo_2021}, an iterative method was proposed to optimise the energy consumption of low-power ground devices served by multiple fixed-wing \acp{uav}, by adjusting the \ac{uav} trajectories to allow for low-power data transmission from the ground devices. Note that this work focused exclusively on the energy consumption of ground devices, and not the \acp{uav} themselves. Energy consumption of multiple fixed-wing \acp{uav} was optimised in \cite{PeterHong_2020}, where the authors presented a trajectory planning heuristic to minimise the energy consumption and increase flight time of a network of fixed-wing \acp{uav} that relay data via multi-hop links. The authors demonstrate that increasing the radius of a \ac{uav}'s circular orbit can reduce the energy consumption for that \ac{uav}. The work in \cite{Shriwastav_2020} proposed deployments that maximise coverage while improving resilience against \ac{uav} failure, using circle-packing theory. Note that as \cite{Kuo_2021}, \cite{PeterHong_2020}, \cite{Shriwastav_2020} focus on basic network coverage scenarios, they do not consider data throughput in their system models, and their optimisation algorithms do not optimise the amount of data that can be delivered across the \ac{uav}-user access link. 



\ac{rl} has been widely investigated to address the energy management challenge in networks of rotary-wing \ac{uav} access points \cite{Liu2019UAVqlearning},~\cite{omoniwa_2021}. In~\cite{Liu2019UAVqlearning}, a cluster-based Q-learning approach was used to optimise the 3D trajectory of multiple \acp{uav} in the network. A distributed multi-agent Q-learning approach was proposed in our prior work~\cite{omoniwa_2021} which took into account the interference from neighbouring \acp{uav}. However, the tabular Q-learning technique presented in \cite{Liu2019UAVqlearning} and~\cite{omoniwa_2021} may not scale with larger state-spaces, thereby requiring deep RL architectures such as those used in our recent work~\cite{Galkin_2022}. It is important to note that rotary-wing \acp{uav} have the advantage of being able to hover in place and move in any direction; this allows for more straightforward trajectory optimisation scenarios. Fixed-wing \acp{uav}, meanwhile, must fly with a certain minimum velocity at all times to remain airborne, and are incapable of making sharp trajectory changes to the same extent as rotary-wing \acp{uav}. For this reason, optimising the system's \ac{ee} in a network of fixed-wing \acp{uav} is more complex, and requires an entirely different approach.



Motivated by these findings, in this paper we focus on optimising the \ac{ee} of multiple fixed-wing \ac{uav} access points, by balancing both the \ac{uav} energy consumption and delivered user throughput. We propose a decentralised, multi-agent \ac{rl} approach, where each \ac{uav} is equipped with a \ac{ddqn} agent which can adjust the \ac{uav} flight trajectory. The \ac{uav} agent communicates with the agents of its neighbouring \acp{uav}, and uses the obtained information to adjust its trajectory in a way that optimises the overall \ac{ee} for itself and its neighbours. Unlike prior works, we consider an interference-limited scenario where nearby \acp{uav} can negatively affect user throughput, which allows for a more in-depth optimisation problem. We compare our proposed multi-agent approach against a series of heuristic trajectory planning strategies, under varying sizes of the deployed \ac{uav} network. This allows us to demonstrate that our solution is able to optimise the system-wide \ac{ee}, while still delivering high-quality data throughput to users on the ground.  To our knowledge, we are the first to investigate the problem of optimising the \ac{ee} of fixed-wing \ac{uav} access points in an interference-limited multi-\ac{uav} network scenario, using tools from deep \ac{rl}.

\vspace{-2mm}
\section{System Model} 
\label{sec:SystemModel}

\begin{figure}[!t]
\centering
\includegraphics[width=0.4\textwidth]{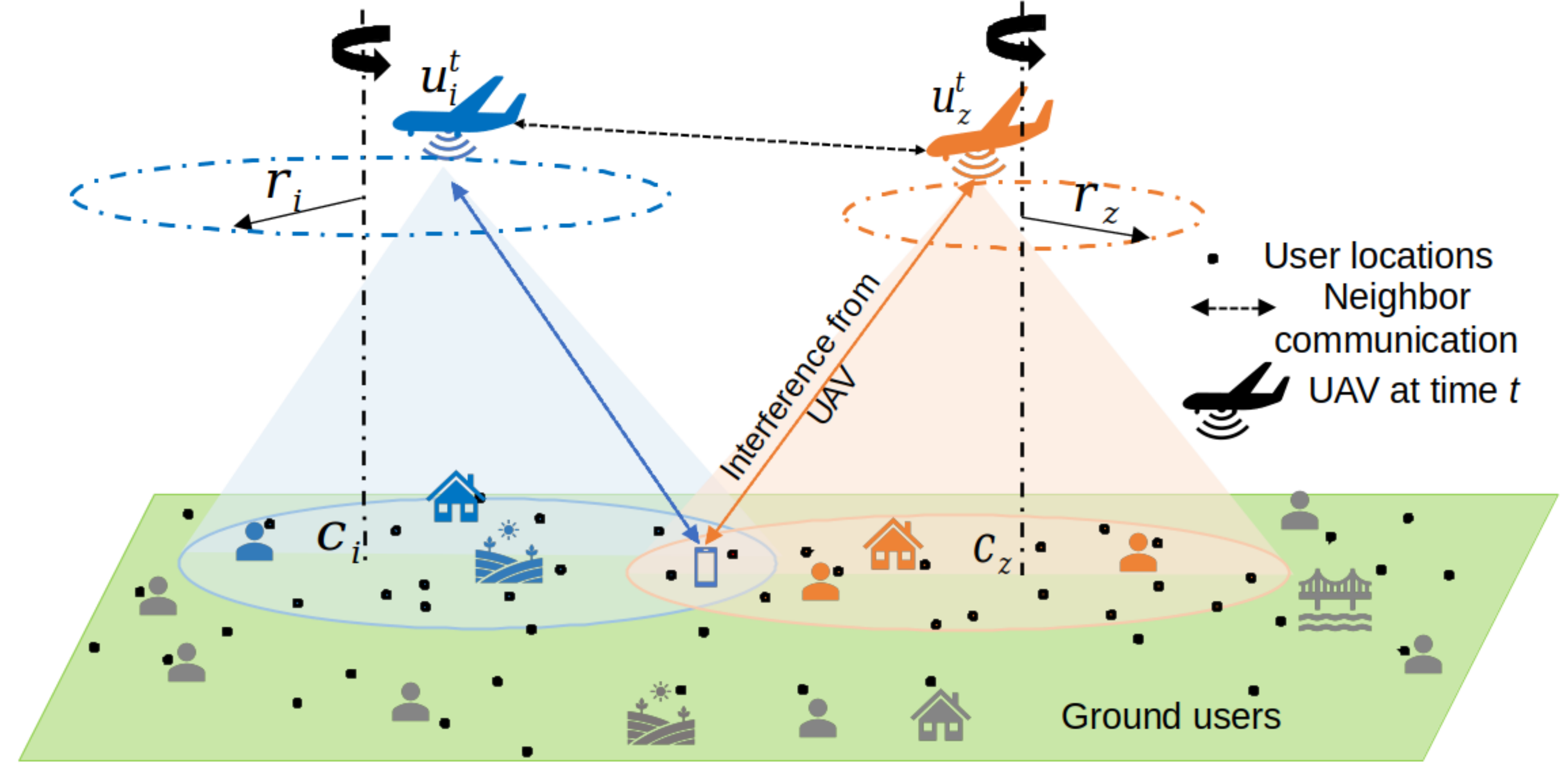}
\caption{Fixed-wing UAVs providing wireless coverage to ground devices.}
\label{scenario}
\vspace{-5mm}
\end{figure}

We consider a suburban or rural environment. In this environment there exist a number of users that require cellular service over a certain period of time T. We discretise this time into $T$ timesteps of duration $\tau$. We denote the set of the users as $\Phi = \{y_1,y_2,y_3....\} \in \Rb^2$. These users are assumed to be distributed as a homogeneous \ac{ppp}. Similar to the related work, we assume the users are static in the environment; the scenario of mobile users is left for a future work.

We assume that a set of \acp{uav} $\mathcal{U} = \{u_1^t,u_2^t,...,u_U^t\} \in \Rb^2$ are deployed by a centralised entity such as a network operator to deliver service to the ground users, with $u_i^t$ denoting the horizontal coordinates of the \ac{uav} $i$ at timestep $t$. This entity has high-level knowledge of the environment, such as the user locations or knowledge about certain points of interest. It uses this information to assign service areas to the \acp{uav}, with the service center-points $\mathcal{C} = \{c_1,c_2,..,c_U\} \in \Rb^2$, where $c_i$ corresponds to the \ac{uav} $i$ at location $u_i^t$. The \acp{uav} are assumed to have a fixed-wing design and cannot hover in place, and so must orbit around their service areas. At timestep $t$, the \ac{uav} $i$ orbits around its center-point $c_i$ with a velocity $v_i^t$, at a height above ground $h_i^t$ and a radius $r_i^t = ||c_i-u_i^t||$. 

The \acp{uav} have downtilted antennas with cone-shaped coverage patterns. The antenna gain (in dB) from \ac{uav} $i$ to user $j$ at timestep $t$ is given as

\vspace{-3mm}
\begin{equation}
\mu(y_j,u_i^t) = -\min\left(20, 12\left(\frac{\arctan(||y_j-u_i^t||/h_i^t)}{\eta}\right)^2\right),
\label{eq:SINR}
\end{equation}

where $\eta$ denotes the \ac{uav} antenna half-power beamwidth. The users are assumed to have omnidirectional antennas with unitary gain.

In each timestep, each user in $\Phi$ associates to the \ac{uav} which provides it with the strongest received signal. If at time $t$ the user $j$ is associated with \ac{uav} $i$, the \ac{sinr} observed by the user is given as 

\vspace{-3mm}
\begin{equation}
\sinr_{i,j}^t = \frac{p c \mu(y_j,u_i^t) ((||y_j-u_i^t||)^2+(h_i^t)^2)^{-\alpha/2}}{\hspace{-3mm}\sum\limits_{k \in \mathcal{U} \setminus {i}}\hspace{-3mm}p c \mu(y_j,u_k^t) ((||y_j-u_k^t||)^2+(h_k^t)^2)^{-\alpha/2}+\sigma^2} ,
\label{eq:SINR}
\end{equation}

\noindent
 where $p$ is the \ac{uav} transmit power, $c$ is near-field pathloss, $\alpha$ is the pathloss exponent, and $\sigma^2$ is the noise power. Note that, as we are considering \acp{uav} operating in the sky above suburban or rural areas, we assume the wireless channel between the \acp{uav} and the users is always \ac{los}.
 
 Whenever a user $j$ associates with a \ac{uav} $i$ we assume it is allocated spectral resources of $B$ bandwidth. As such, the data throughput for user $j$ from \ac{uav} $i$ at timestep $t$ is given by the Shannon bound as 
 
 \begin{equation}
R_{i,j}^t = \tau B\log_2(1+\sinr_{i,j}^t).\label{eq:throughput}
 \end{equation}
 
Naturally, $R_{i,j}^t = 0$ for all the \acp{uav} in $\mathcal{U}$ other than $i$.
 
At each timestep, the \ac{uav} $i$ consumes a certain amount of energy $E_i^t$ given in \cite{Zeng_2017} as 

\begin{equation}
E_i^t = \tau\Bigg(\left(c_1+\frac{c_2}{(g r_i^t)^2}\right)(v_i^t)^3 + \frac{c_2}{v_i^t}\Bigg),
\label{eq:energyconsumption}
\end{equation}

where $c_1$ and $c_2$ are parameters related to the \ac{uav} aerodynamic design \cite{Zeng_2017}  and $g$ is the gravitational constant. The \ac{uav} communication equipment will also consume energy, but we assume that this energy consumption is significantly smaller than the energy associated with flight \cite{Zeng_2017}, and therefore we only consider the latter in this work.
Given the above equation, the velocity which minimises the energy consumption for the given turn radius is derived in \cite{Zeng_2017} as:

\begin{equation}
v^*(r) = \Bigg(\frac{c_2}{3(c_1+c_2/(gr)^2)}\Bigg)^{1/4}.
\label{eq:optV}
\end{equation}

Note that when the \ac{uav} flies at the optimum velocity for its given turn radius, the energy consumption monotonically decreases with increasing turn radius. The \ac{ee} of \ac{uav} $i$ at timestep $t$ is defined as 

\vspace{-2mm} 
\begin{equation}
EE_i^t = \frac{\sum\limits_{j \in \Phi} R_{i,j}^t}{E_i^t}.
\end{equation}
\vspace{-2mm} 

It follows that the total \ac{ee} of the network across the entire period T is given as 

\vspace{-2mm}
\begin{equation}
EE_{tot} = \frac{\sum\limits_{t=1}^T\sum\limits_{i \in \mathcal{U}}\sum\limits_{j \in \Phi} R_{i,j}^t}{\sum\limits_{t=1}^T\sum\limits_{i \in \mathcal{U}}E_i^t} .
\end{equation}
\vspace{-2mm}

\section{Energy Efficiency Optimisation}
\subsection{Problem Statement}
When they are deployed by their central entity, the \acp{uav} are assumed to travel to their assigned service areas and enter into a circular orbit around the center-points at the minimum possible turn radius $r_{\min}$, while delivering service to the users. Once all of the \acp{uav} arrange themselves in this manner (at the beginning of the episode when $t=1$), each \ac{uav} begins to adjust its trajectory around its center-point. In this paper we focus on optimising the \acp{uav} once they arrive at their service areas; optimising \ac{uav} trajectories during the initial flight to the center-points is left for a future work. The optimisation problem for the \ac{uav} network is given as 

\vspace{-3mm}
\begin{subequations}
\begin{align}
&\underset{(\mathbf{r_1},...\mathbf{r_T}),(\mathbf{h_1},...,\mathbf{h_T})}{\max}  EE_{tot} \\
\text{s.t.} \quad &r_{\min} \leq r_i^t \leq r_{\max},\quad \forall i, t \\
&h_{\min} \leq h_i^t \leq h_{\max},\quad \forall i, t
\end{align}
\end{subequations}
\vspace{-3mm} 

where $\mathbf{r_t} = (r_1^t,r_2^t,...,r_U^t)$ and $\mathbf{h_t} = (h_1^t,h_2^t,...,h_U^t)$ are vectors of the \ac{uav} radii and heights at timestep $t$, respectively. In other words, we wish to maximise the total \ac{ee} of the \ac{uav} deployment across the entire episode, by optimising the \ac{uav} radii and heights at each timestep. The two constraints ensure that the \ac{uav} heights and radii stay within set bounds. As the user locations are assumed to be static the center-points $\mathcal{C}$ are assumed to not change. As such, the \acp{uav} can optimise their \ac{ee} by adjusting their radii, heights, and velocities. Note that the relationship between the velocity and turn radius is defined as in \Eq{optV}; we assume that the \acp{uav} are aware of this relationship and so always adopt the corresponding velocity for their given turn radius.

\vspace{-2mm}
\subsection{Solution}
To address the optimisation problem defined above, we propose applying a decentralised multi-agent \ac{rl} solution. Each \ac{uav} is equipped with an agent which can gather information about the environment and adjust the \ac{uav} trajectory. 

For a \ac{uav} $i$, let us denote the set of neighbour \acp{uav} as belonging to the set $\mathcal{U}_i \subset \mathcal{U}$. The set $\mathcal{U}_i$ contains the six \acp{uav} whose center-points are closest to the center-point of $i$, or all of the \acp{uav} other than $i$ if there are less than six \acp{uav} deployed in total. The reason we consider six \acp{uav} is due to the assumption that when the \acp{uav} are deployed, their service areas will roughly correspond to the shape of a hexagonal lattice, with each \ac{uav} being surrounded by a ring of six other \acp{uav} that have the most impact on its throughput via interference. 

In each timestep, each \ac{uav} agent will communicate with the agents of its neighbouring \acp{uav} and share information. The agents will then use this information to sequentially choose which action to take next. When an agent chooses an action for the current timestep, it broadcasts its choice to its neighbours, so that those neighbours who have not yet chosen an action in the current timestep can use that information for their own decision-making.  When all of the agents have decided what action to take next, they will carry out their respective actions simultaneously, transmit their data to the users and consume the corresponding amount of energy each. 


\begin{figure}[!t]
\centering
\vspace{-2mm}
\includegraphics[width=0.45\textwidth]{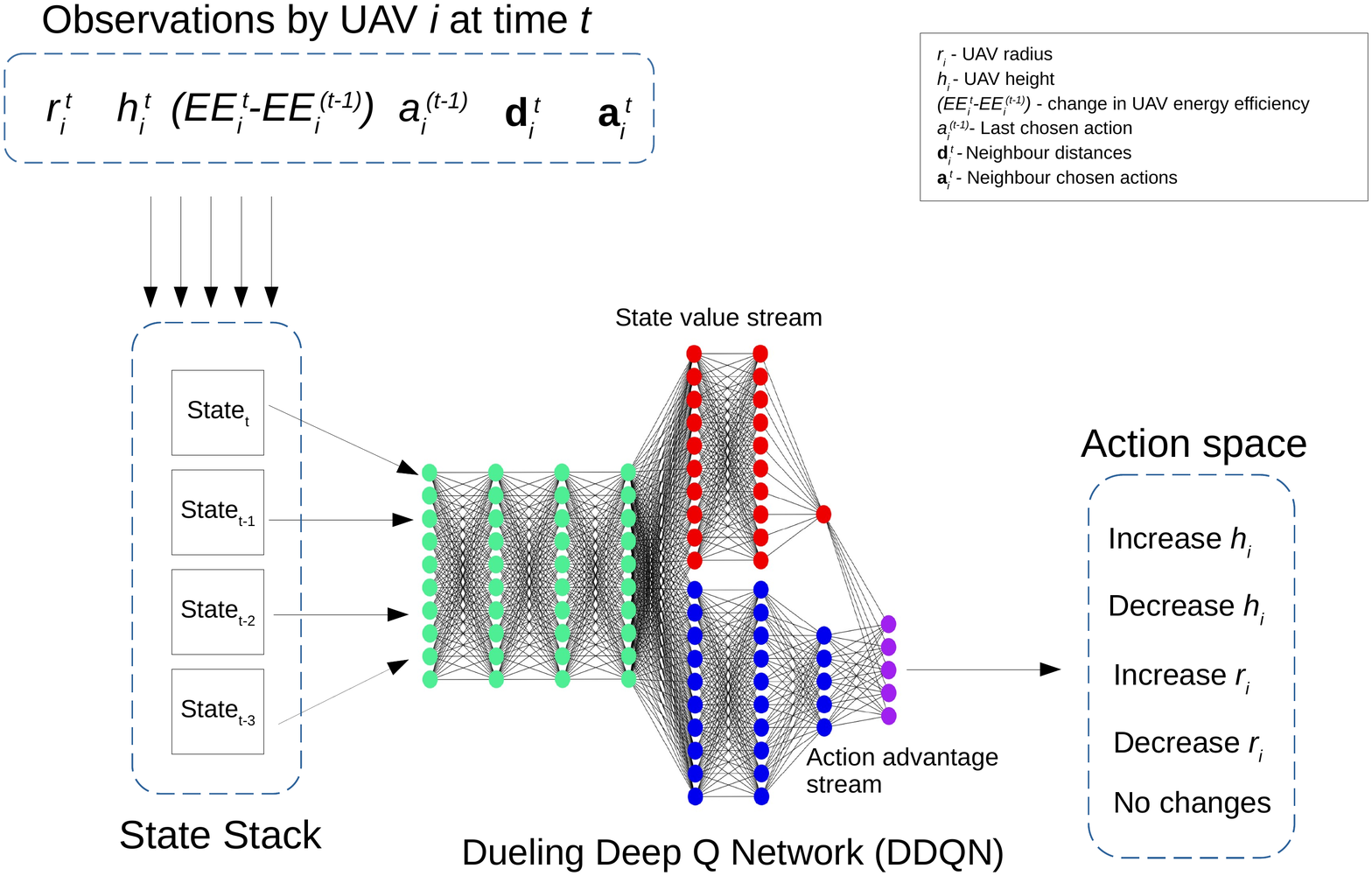}
\vspace{-4mm}
\caption{\centering Our proposed \ac{ddqn} solution. Note that the number of neurons per \ac{ddqn} layer is reduced in the diagram for ease of readability.}
\label{fig:network}
\vspace{-3mm}
\end{figure}

We propose using a \ac{ddqn} for each \ac{uav} agent, as depicted in \Fig{network}. The \ac{ddqn} takes in a vector of observations as an input, chooses an action to take, and observes a reward.  These are defined below

\subsubsection{States}
At timestep $t$ the state $\textbf{s}_t$ for \ac{uav} $i$ is a vector which contains:
\begin{itemize}
    \item the current \ac{uav} turn radius $r_i^t$.
    \item its height above ground $h_i^t$.
    \item the latest change in the \ac{uav}'s \ac{ee} $EE_i^t-EE_i^{(t-1)}$.
    \item the action taken by $i$ in the previous timestep, $\textit{a}_i^{(t-1)}$.
    \item the horizontal distances of the neighbour \acp{uav} in the set $\mathcal{U}_i$ to the center-point $c_i$ at timestep $t$. We denote these distances as a vector $\textbf{d}_i^t = (||u_1^t-c_i||,...,||u_j^t-c_i|| ), \forall u_j \in \mathcal{U}_i$
    \item the index of the actions selected by the neighbour \acp{uav} that have already chosen an action to take in this timestep, denoted as the vector $\textbf{a}_i^t = (a_1^t,...,a_j^t), \forall u_j \in \mathcal{U}_i$.
\end{itemize}

After the state vector $\textbf{s}_t$ is formed, it is passed into a state stack and combined with the state vectors from the previous three timesteps. This combined vector is then passed into the \ac{ddqn}. By combining the state vector for multiple timesteps together the agent is able to observe not just the environmental state at that moment, but also the changing dynamics over time based on the chosen actions.

\subsubsection{Actions}
Each \ac{uav} can take one of five actions at timestep $t$: increase/decrease the radius by increment $r_{inc}$, increase/decrease the height by $h_{inc}$, or keep radius and height the same and continue traveling on its current trajectory. If the \ac{uav} changes its radius, it will also adjust its velocity to the one that minimises energy consumption as per \Eq{optV}.

\subsubsection{Rewards}
When all of the agents have chosen which action to take next, all \ac{uav} locations are updated for timestep $t+1$, and the \acp{uav} observe their individual throughput and energy consumption. This performance data is shared among neighbour \acp{uav}, and the reward for the $i$-th agent at timestep $t+1$ is calculated as:

\vspace{-2mm} 
\begin{equation}
\rho_i^{(t+1)} = \frac{\sum\limits_{k \in \mathcal{U}_i\cup {i}}\sum\limits_{j \in \Phi} R_{k,j}^{(t+1)}}{\sum\limits_{k \in \mathcal{U}_i\cup {i}}E_k^{(t+1)}} - \frac{\sum\limits_{k \in \mathcal{U}_i\cup {i}}\sum\limits_{j \in \Phi} R_{k,j}^{t}}{\sum\limits_{k \in \mathcal{U}_i\cup {i}}E_k^{t}}
\end{equation}
\vspace{-2mm} 

In other words, the reward for the $i$-th agent is the change in the total \ac{ee} for itself and its neighbours. We assume that by maximising this \ac{ee} based on local observation data across the different timesteps we will also maximise the total episode \ac{ee} $EE_{tot}$.

\subsubsection{\ac{ddqn} implementation}
The \ac{ddqn} is a neural network which consists of several dense, feedforward neuron layers. There are 4 layers in the main stream, which then split off into a State Value stream and an Action Advantage stream with three layers each, as per the \ac{ddqn} design \cite{Wang_2016}. These two streams finally combine together in a combination layer. The combination neuron layer and the final layer of the Action Advantage stream have five neurons each, one for each of the five actions. the final layer in the Value stream has a single neuron. The rest of the layers in the \ac{ddqn} have 64 neurons. All of the layers with the exception of the combination layer use a reLU activation function.

\subsubsection{\ac{ddqn} training}
\ac{ddqn}, like all \ac{rl} processes, is trained in an online manner. The agent takes actions in the environment, observes the results, and gradually learns how to act to maximise its reward. After an agent takes an action, it stores the previous state stack, the action taken, the received reward, and the new state stack in a database called a replay buffer. After enough timesteps, the replay buffer will be sufficiently populated to use for training. Using uniform random sampling, a batch of entries will be selected from the replay buffer and used to update the weights of the \ac{ddqn}, following the methodology described in \cite[Chapter 4]{Francois-Lavet_2018}.

To ensure that the agent is able to explore and populate the replay buffer with a wide range of observations, we use an $\epsilon$-greedy policy, where the agent will take actions randomly with probability $\epsilon$, and will take an action based on the \ac{ddqn} output otherwise. $\epsilon$ has a value of $1$ initially, and this is decayed by a certain factor at the end of every timestep. This allows the agent to explore in the beginning of the training process when the \ac{ddqn} is untrained, while having it rely on the \ac{ddqn} more and more as it gets better at making decisions.

\vspace{-1mm}
\section{Evaluation Scenarios and Results}
\vspace{-1mm}

\begin{table}[t]
\vspace{-3mm}
\begin{center}
\caption{Numerical Result Parameters}
\vspace{-3mm} 
\begin{tabular}{|c|c|} 
 \hline
 Parameter & Value\\ 
 \hline
 Carrier Frequency & \unit[2]{GHz} \\
 pathloss exponent $\alpha$ & 2.1 \\
\ac{uav} transmit power $p$ & \unit[1]{W} \\
\ac{uav} half-power beamwidth $\eta$ &  \unit[30]{deg.} \\
\ac{uav} mass & \unit[10]{kg} \\
aerodynamic parameter $c_1$ & $9.26\cdot 10^{-4}$ \cite{Zeng_2017} \\
aerodynamic parameter $c_2$ & 2250 \cite{Zeng_2017} \\
 Near-field pathloss $c$ & \unit[-38.4]{dB} \cite{Elshaer_2016}\\
 Noise power $\sigma^2$ & \unit[$8\cdot10^{-13}$]{W} \cite{Elshaer_2016}\\ 
User density & \unit[10]{$/\text{km}^2$} \\
UAV density & \unit[0.2]{$/\text{km}^2$} \\
\ac{uav} height bounds $h_{\min}$, $h_{\max}$ & \unit[20]{m}, \unit[300]{m} \\
\ac{uav} radius bounds $r_{\min}$, $r_{\max}$ & \unit[50]{m}, \unit[1000]{m} \\
\ac{uav} height increment $h_{inc}$,& \unit[5]{m}\\
\ac{uav} radius increment $r_{inc}$ & \unit[10]{m} \\
  MC trials & 500\\
 Episodes per MC trial & 1\\
 Timesteps per episode $T$ & 250\\
 Timestep duration $\tau$ & 2 seconds \\
 Q-value discount factor & 0.1 \\
 Learning rate & 0.00005 \\
 Initial epsilon value $\epsilon$ & 1 \\
 Epsilon decay value & 0.99995 \\
 Minimum epsilon value & 0.001\\
 Replay memory size & 5000 entries\\
 Replay batch size & 1000 \\
 \hline
\end{tabular}
 \label{tab:table}
\end{center}
\vspace{-5mm}
\end{table}

\begin{figure}[!t]
\centering
\vspace{-2mm}
\includegraphics[width=0.30\textwidth]{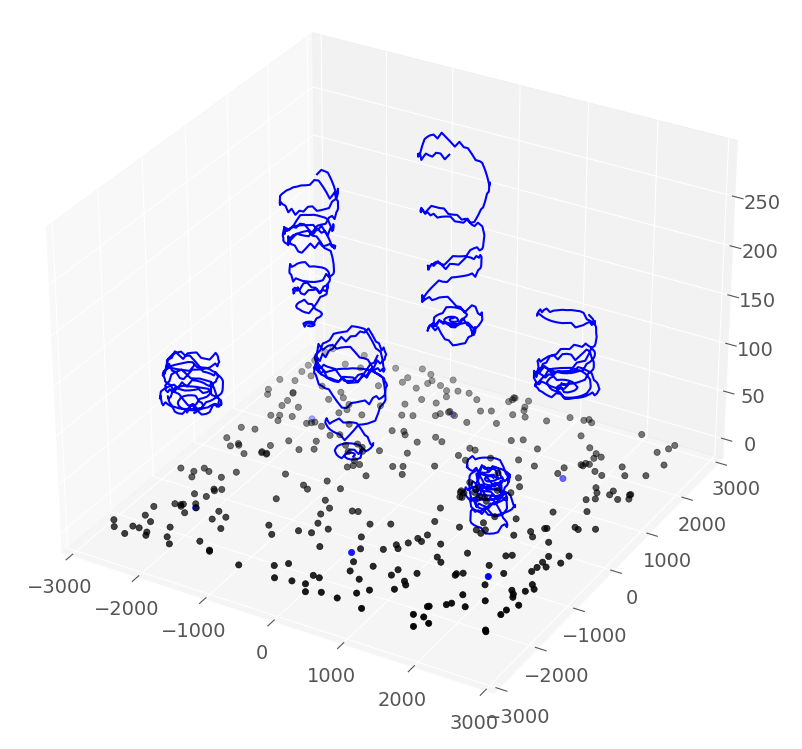}
\vspace{-4mm} 
\caption{3D orbits of six deployed \acp{uav} following our proposed algorithm.}
\vspace{-3mm} 
\label{fig:orbits}
\vspace{-3mm} 
\end{figure}

In this section we evaluate the performance of our proposed algorithm using simulations, across deployments of 2-20 \acp{uav}. We generate a number of \ac{mc} trials, with a random distribution of users in each trial distributed in a square simulation area of a certain size. To preserve the \ac{uav} and user densities, we scale up the size of the simulation area with the increasing number of deployed \acp{uav}, with simulation areas in the range of \unit[10-100]{$\text{km}^2$}, and user numbers in the order of 100-1000. The center-points are found for a given user distribution and number of deployed \acp{uav} using the K-Means Clustering algorithm \cite{Galkin_2016}. The \acp{uav} are deployed in a circular orbit around their center-points, at an initial height of $\unit[100]{m}$ and the minimum possible radius $r_{\min}$. The reason we intialise the \acp{uav} to fly at the minimum radius is that this ensures they are as close as possible to the center-points, and therefore can achieve the highest throughput to the users, but at the worst possible energy consumption. The training episode will then begin, and the \acp{uav} act to optimise their flight trajectories with respect to the system \ac{ee}. The numerical result parameter values are given in the table. \Fig{orbits} shows an example of the \ac{uav} orbits for one of the \ac{mc} trials.

As the problem of optimising the \ac{ee} of fixed-wing \ac{uav} access points in an interference-limited environment is quite novel, there is a lack of available benchmarks against which to compare the performance of our multi-agent solution. As such, we propose several heuristics for the results comparison:
\begin{itemize}
    \item Minimum radius orbit. For this heuristic the \acp{uav} remain at their initial heights and turn radii, and follow their simple circular orbits for the entire episode.
    \item Quadcopter-style \ac{uav} hovering above the center-points. For this heuristic, the \acp{uav} are assumed to have a quadcopter design, and are assumed to hover in place directly over their center-points throughout the episode, as in \cite{Galkin_2016}. This represents the type of \ac{uav} access point network most commonly investigated in the literature, due to their flexible mobility and hovering capability. To model their energy consumption, we use Eq. (2) in \cite{Kingry_2018}.
    \item Bounded random walk. In this heuristic, the \acp{uav} choose their actions randomly in each timestep, provided the heights and radii remain within the permitted bounds. As our \ac{rl} solution relies on $\epsilon$-greedy exploration during training, this benchmark represents the performance bound of our solution before it is trained.
    \item Energy-saving orbit. In this heuristic, all of the \acp{uav} fly in simple orbits at a radius equal to half the minimum distance between two center-points. This represents the maximum orbits that the \acp{uav} can fly at without any overlaps occurring, and is aimed at minimising the energy consumption.
\end{itemize}

\begin{figure}[!t]
\centering
\includegraphics[width=0.45\textwidth]{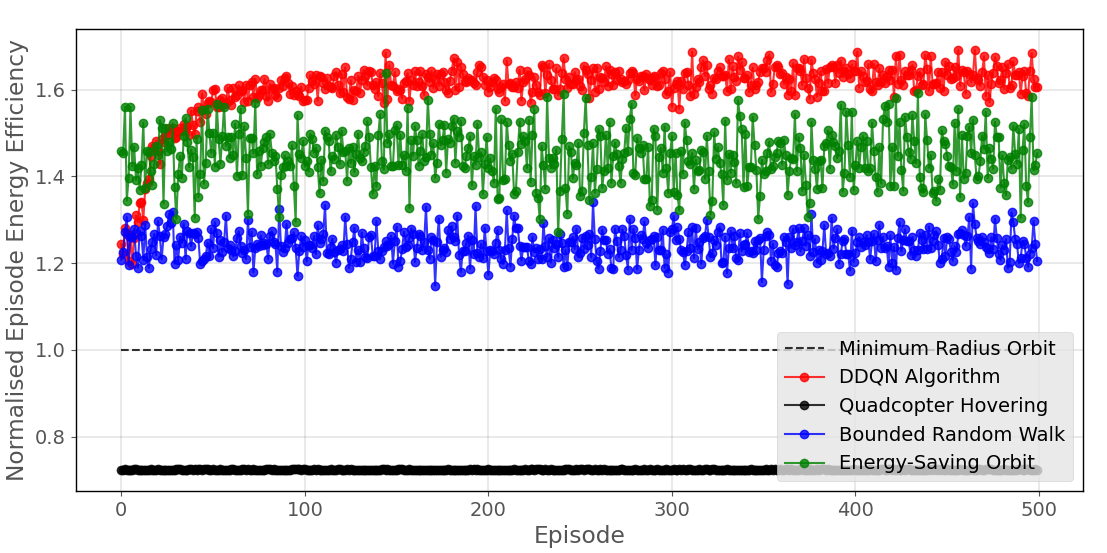}
\vspace{-4mm} 
\caption{\centering Normalised \ac{ee} performance of the solutions over the range of episodes, given 10 \acp{uav}.}
\vspace{-3mm} 
\label{fig:training}
\vspace{-3mm} 
\end{figure}

\Fig{training} shows how our algorithm performs as it is trained over a number of episodes. To improve the clarity of our results, for each episode the overall system performance results of the algorithm and the heuristics are normalised with respect to the performance of the minimum radius orbit heuristic. This allows us to display the results as ratios of the minimum radius baseline. We observe that in the beginning of the training process, our algorithm (shown in red) relies on random action exploration according to the $\epsilon$-greedy policy, and as such gives performance comparable to the bounded random walk heuristic (shown in blue). However, after approximately 100 episodes the algorithm reaches a point where it can reliably outperform all of the heuristics that we test against. Note that each episode corresponds to its own \ac{mc} trial, which means that for each episode we randomise the positions of the ground users, and recalculate the center-points for the \acp{uav}. As a result, the performance fluctuates from episode to episode. It is important to note that our algorithm is able to deal with this randomness, and is able to give good performance even when applied to a "new" environment with user distributions that it had not been trained on previously.

\begin{figure}[!t]
\centering
\includegraphics[width=0.45\textwidth]{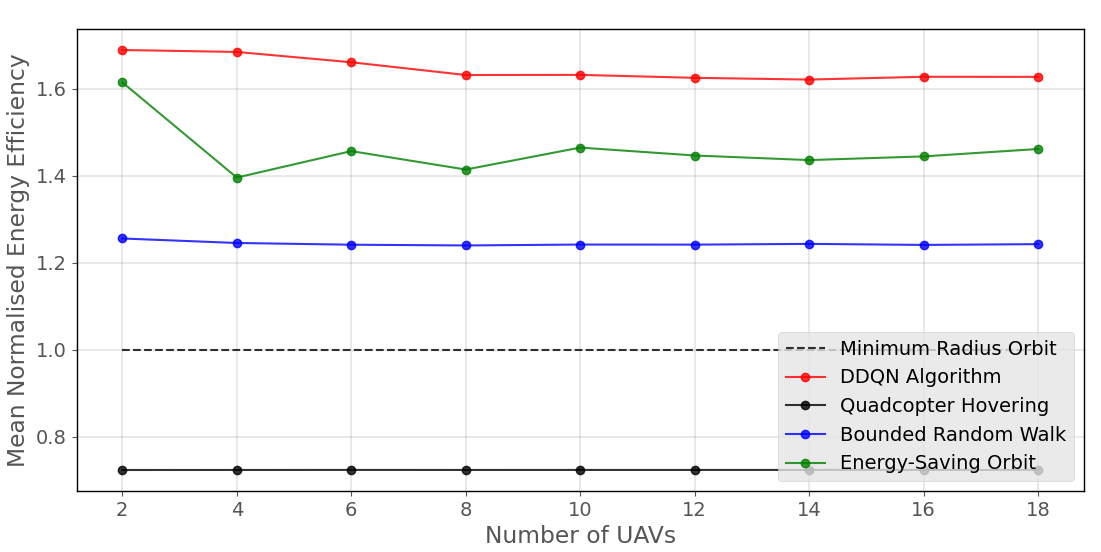}
\vspace{-4mm} 
\caption{\centering Normalised \ac{ee} performance of the solutions for different \ac{uav} numbers.}
\vspace{-3mm} 
\label{fig:EE}
\end{figure}

\begin{figure}[t]
\centering
\includegraphics[width=0.45\textwidth]{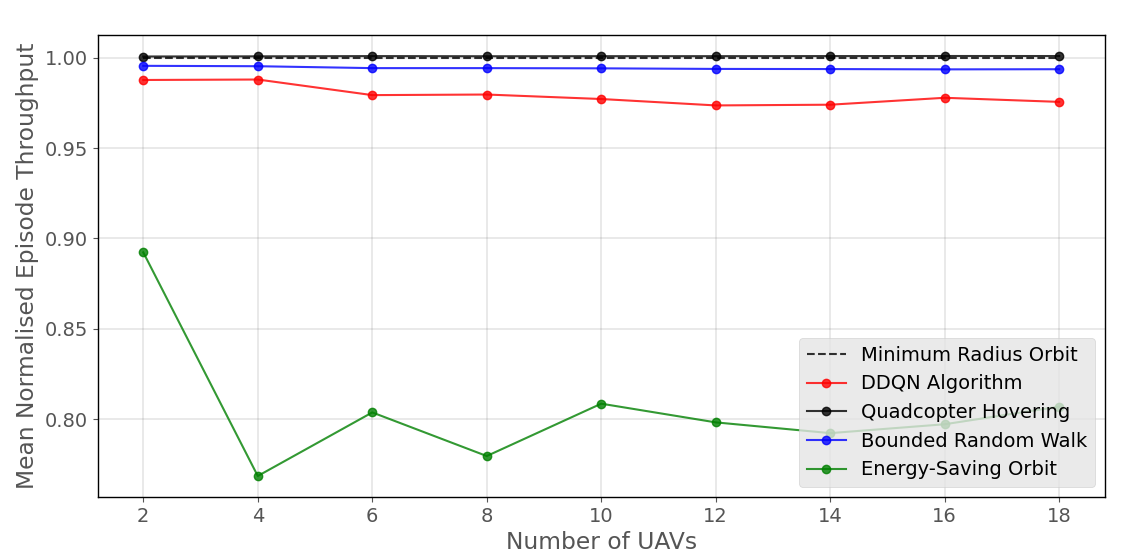}
\vspace{-4mm} 
\caption{\centering Normalised system throughput performance of the solutions for different \ac{uav} numbers.}
\vspace{-3mm} 
\label{fig:throughput}
\vspace{-3mm} 
\end{figure}

\begin{figure}[t]
\centering
\includegraphics[width=0.45\textwidth]{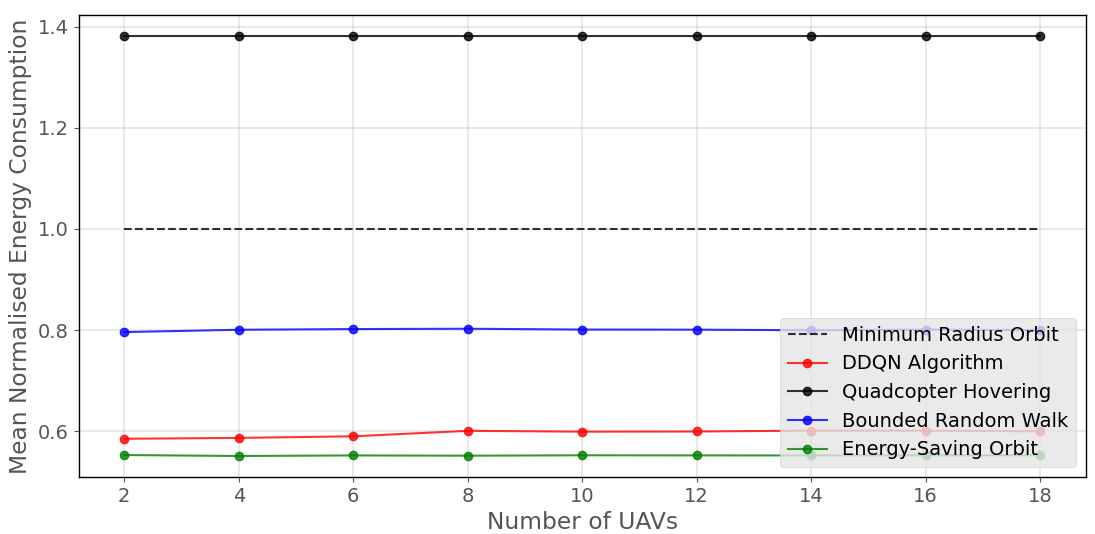}
\vspace{-4mm} 
\caption{\centering Normalised system energy consumption of the solutions for different \ac{uav} numbers.}
\vspace{-3mm} 
\label{fig:energy}
\vspace{-3mm} 
\end{figure}

To verify that our algorithm can scale for varying numbers of deployed \acp{uav}, we generate our numerical results for a range of \ac{uav} fleet sizes. For each fleet size, we rerun our training process for 500 episodes, and record the normalised performance metrics for the 400 episodes after training concludes. The resulting mean values for \ac{ee}, episode throughput, and energy consumption are shown in \Fig{EE} to \Fig{energy}.

Our first observation is that the fixed-wing \ac{uav} deployments, and our proposed algorithm in particular, all significantly outperform the hovering quadcopter heuristic in terms of \ac{ee} across the entire range of \ac{uav} deployments. This is explained by investigating the throughput and energy consumption figures. \Fig{throughput} shows that hovering directly over the center-points instead of maintaining a close orbit around them gives minimal improvement to the resulting user channel throughput. However, \Fig{energy} shows that hovering directly in place consumes substantially more energy than the fixed-wing \acp{uav}, as the hovering \acp{uav} cannot benefit from the effect of aerodynamic lift, and must instead generate enough downward thrust to counter the force of gravity. This is in line with real-world \ac{uav} performance, and once more highlights the importance of using fixed-wing designs in scenarios where the \acp{uav} must stay in the air for extended periods of time.

The figures also show that our algorithm consistently gives the best \ac{ee} performance, compared to the other heuristics. The algorithm is able to delver up to 70\% more \ac{ee} compared to the minimum radius orbit baseline. We also note that increasing the number of \acp{uav} in the network reduces the gains by only approximately 10\%. As expected, the majority of the interference for a \ac{uav} comes from its six closest neighbours, and as such the \acp{uav} can successfully optimise the global performance based only on the local observations from their nearest neighbours.

Looking at \Fig{throughput} we see that optimising the \ac{ee} does result in the \ac{ddqn} algorithm sacrificing some throughput, but this appears to be only in the order of a couple of percent compared to the hovering or minimum radius orbit performance. Overall, the \ac{uav} network is able to maintain consistently high-quality throughput for the ground users.

The real performance gains arise from the energy consumption, shown in \Fig{energy}. By intelligently adjusting their flight trajectories, the \acp{uav} with the \ac{ddqn} algorithm are able to intelligently increase their orbit radii and reduce their energy consumption by approximately 40\% compared to the baseline. Note that the bounded random walk heuristic also manages to reduce the energy consumption by randomly drifting away from the center-points. 

Finally, we observe that the energy-saving orbit heuristic is able to give good performance, but only for very small numbers of \acp{uav} when the aggregate interference is low. As the \acp{uav} fly far away from their center-points they are significantly more vulnerable to interference than the other deployment strategies which fly closer to the center-points. Note that there is some fluctuation in the performance for 4-10 drone scenarios, we attribute this to the variable orbit distances that arise from placing wide, non-overlapping \ac{uav} orbits in square simulation areas.

\vspace{-2mm} 

\section{Conclusion}

In this work we have demonstrated a multi-agent deep \ac{rl} approach to optimise the \ac{ee} of a fleet of fixed-wing \ac{uav} access points that are deployed to provide service to ground users, under interference-limited channels. We demonstrated that our proposed solution can significantly improve the system \ac{ee} based on local state observations and communication between neighbouring \acp{uav}, while still ensuring that ground users receive high-quality service. In future works, we will extend our study to consider scenarios where users move through the environment, such as high-speed road users. For these scenarios, we will investigate how to optimise the flights of the \acp{uav} to accommodate multiple types of users. In this work we have focused on the ground user access link; in future works, we may also optimise around the wireless backhaul of the \acp{uav}.

\bibliographystyle{./IEEEtran}
\vspace{-2mm} 

\bibliography{./IEEEabrv,./IEEEfull} 

\end{document}